\documentclass[cup6a]{cupconf}
\input psfig.sty
\usepackage{psfig}
\usepackage{epsf}
\usepackage{graphicx}

\title[Dust, Metals and DIBs in Damped Lyman Alpha Systems]
      {Dust, Metals and Diffuse Interstellar Bands in Damped Lyman Alpha Systems}
\author{Sara L. Ellison}
\affiliation{Department of Physics \& Astronomy, University of Victoria,
3800 Finnerty Rd., Victoria, BC, Canada}

\date{}

\begin{document}

\pagenumbering{roman}
\maketitle

\begin{abstract}
Although damped Lyman alpha (DLA) systems are usually considered metal-poor,
it has been suggested that this could be due to observational bias against
metal-enriched absorbers.
I review recent surveys to quantify the particular issue of dust
obscuration bias and demonstrate that there is currently no compelling
observational evidence to support a widespread effect due to extinction.
On the other hand, a small sub-set of DLAs may be metal-rich and
I review some recent observations of these metal-rich absorbers and
the detection of diffuse interstellar bands in one DLA at $z \sim 0.5$.
\end{abstract}

\section{Introduction}

At this conference on the metal-rich universe, a talk on damped Lyman
alpha (DLA) systems may seem misplaced, since there is now a considerable
literature on these absorption selected galaxies that describes
them as metal-poor (e.g. Pettini 2004 and references therein).  
Faced with the almost 
universally sub-solar
abundances of DLAs over all redshifts (the average metallicity of DLAs
is $\sim 1/30 Z_{\odot}$ at $z \sim 3$ and  $\sim 1/10 Z_{\odot}$  
and $z \sim 1$), questions have been
raised concerning DLA selection techniques.  For example, are we
missing a large fraction of metal-rich DLAs due to dust obscuration
bias (Ostriker \& Heisler 1984)?  Or are the bulk of the metals in absorbers
below the canonical DLA column density threshold (P\'eroux et al.
2006)?  In this
contribution, I review the latest observations on the possibility
of selection bias and discuss whether DLAs can ever be considered as
metal-rich.

\section{Are DLA Abundances Biased Due to Dust Obscuration?}

\begin{figure}
\centerline{\psfig{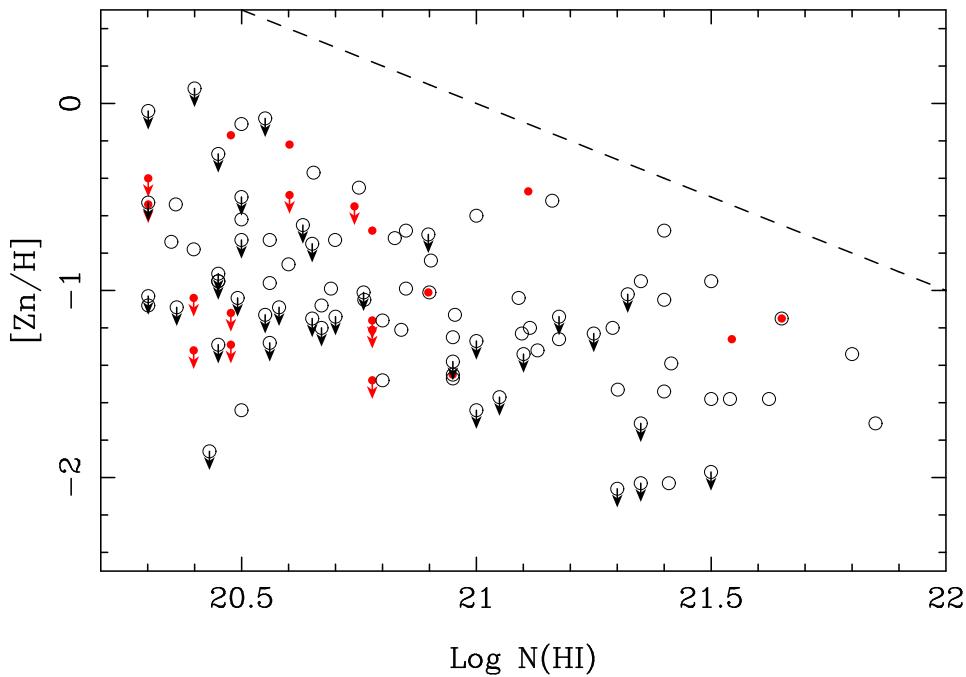}}
\caption{Metallicities from optically selected DLA surveys (open points)
and from the radio-selected CORALS survey (filled points).  The dashed
line shows the `dust filter' of Prantzos \& Boissier (2000), although
this is inconsistent with limits on DLA reddening (Ellison, Hall
\& Lira 2005). Figure adapted from Akerman et al. (2005).}
\end{figure}

\begin{figure}
\centerline{\psfig{file=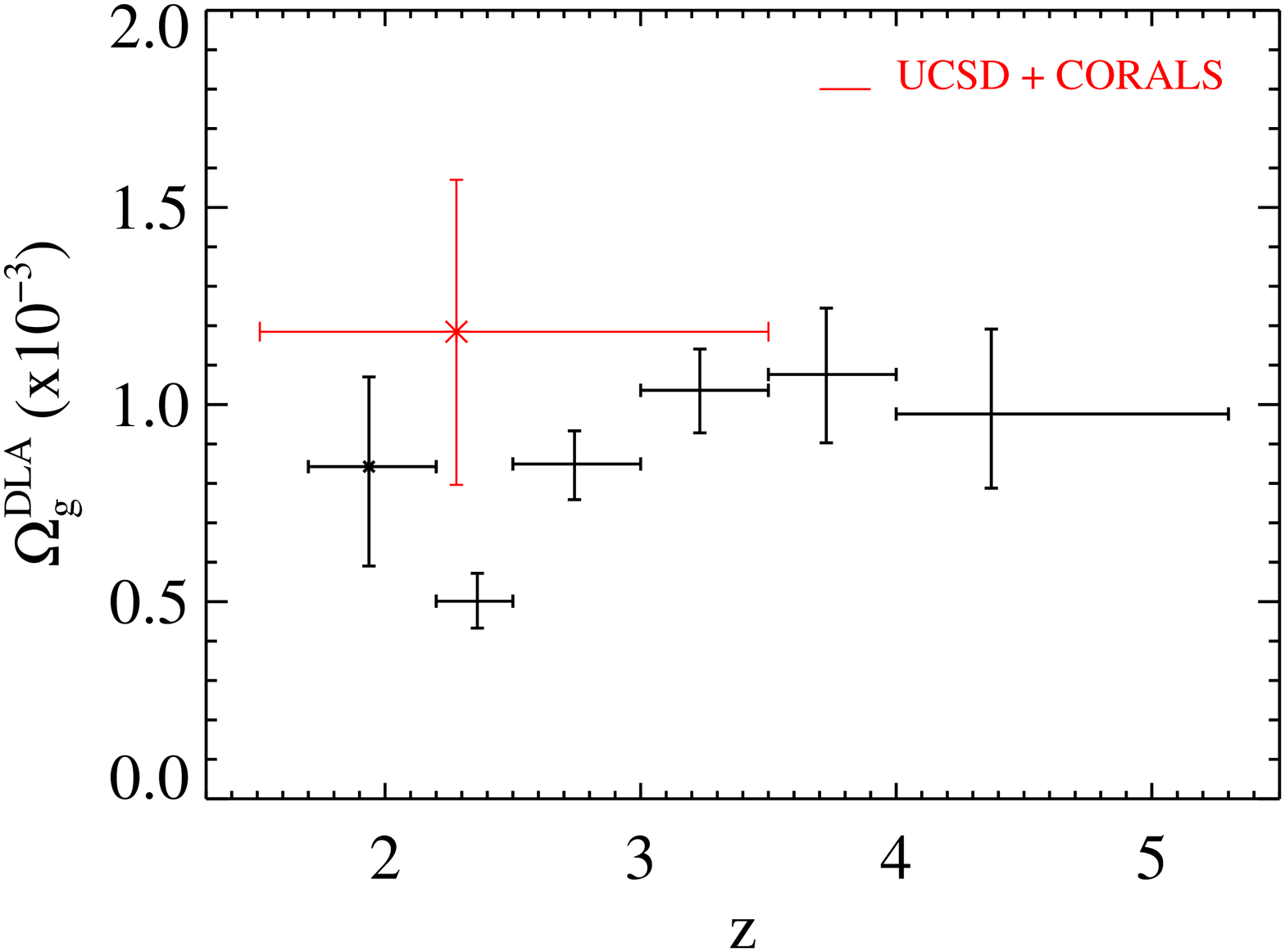,width=5in,angle=90}}
\caption{Mass density of neutral gas in DLAs from SDSS (points shown as
a function of redshift) and radio-selected surveys (cross).
Figure adapted from Jorgenson et al. (2006).}
\end{figure}

DLA identification relies on a bright background QSO on whose continuum
the strong Lyman $\alpha$ and metal line species can imprint their
absorption signature.  If an intervening galaxy is rich in dust and
metals, the very background QSO on which we rely will appear fainter
and redder and may `drop out' of traditional quasar surveys.  This idea
has been around for more than 20 years (Ostriker \& Heisler 1984;
Fall \& Pei 1993)
and it has been suggested that extinction bias may `hide' more
than 70\% of gas at high redshift.  Four pieces of observational 
support are often cited for the dust bias scenario 1) only mild 
redshift evolution in DLA metallicities, 2) low metallicities
compared with emission line galaxies at similar redshifts, 3)
steeper continuum slopes in QSOs with DLAs compared to those
without (Pei, Fall \& Bechtold 1991) and 4) 
an anti-correlation between N(HI) and 
metallicity (Prantzos \& Boissier 2000, see Figure 1).

In order to quantify the impact of dust bias, two surveys based
on radio-selected QSOs have now been completed.  Together,
the Complete Optical and Radio Absorption Line System (CORALS;
Ellison et al. 2001) and the UCSD survey (Jorgenson et al. 2006)
cover a redshift path $\Delta z \sim 100$ for 119 radio-selected
QSOs with very deep (and, in the case of CORALS, complete) optical
follow-up.  Neither survey finds an excess of absorbers either
at high or low (Ellison et al. 2004) redshift and the neutral
gas content is in good agreement (within a factor of 2) with optical
samples, see Figure 2.  More importantly
for the present discussion, Akerman et al (2005) have shown
that the metallicities of DLAs in the CORALS sample are not
significantly higher than optically selected samples and do
not populate the parameter space at high N(HI) and high
metallicity (Figure 1).  However, since metallicities are usually
weighted by the rare high N(HI) absorbers, a larger sample is
required in order to make this result robust.

So, how do we explain the the observational `evidence' in support
of dust bias?  A number of recent papers point to the idea that
although DLAs maybe reservoirs rich in atomic gas, they do not generally
flag the location of the bulk of star-formation and, therefore, 
metals (e.g. Wolfe \& Chen 2006).  Indeed, emission line
spectroscopy of $z \sim 0.5$ galaxies causing absorption line
systems typically have solar abundances (Ellison, Kewley \&
Mallen-Ornelas 2005).  Such observations indicate that high
abundances \textit{can} be found in absorption selected galaxies,
although they maybe confined to smaller regions than the
cross-section of DLA-producing gas.  The cause of the anti-correlation
between N(HI) and metallicity is still under debate.  However,
Ellison, Hall \& Lira (2005) have argued that the very low
values of reddening that are now being determined for DLAs
may soon make dust obscuration an unviable explanation.  An
alternative explanation may be that sightlines that pass through
high column density, high metallicity gas, are simply rare.
Simulations support this idea (e.g. Johansson \& Efstathiou 2006),
showing that the cross-section for such gas at $z \sim 3$
is small.  Finally, concerning the reddening of QSO continua,
this may simply have been a case of a difficult measurement
combined with small number statistics.  Fitting continua to $\sim 1500$
SDSS spectra, Murphy \& Liske (2004) have found reddening
to be very low: E(B$-$V) $<$ 0.02 and a similarly low value
has been found by Ellison, Hall \& Lira (2005) in CORALS
QSOs, E(B$-$V) $<$ 0.04 based on optical-to-IR colours. 
Therefore, regardless of concerns about the modest sample
size of radio-selected QSO surveys such as UCSD and CORALS,
there is no compelling observational evidence to invoke a dust
bias. 
Combined with this revised observational view of dust obscuration, it is
interesting to note that theory is also re-assessing the
effect of extinction.  For example, Trenti \& Stiavelli (2006)
estimate the total gas density in DLAs to be underestimated
by only $\sim 15$\% in optical surveys.

\section{Are All DLAs Low Metallicity?}

Although the majority of DLAs are metal and dust poor, this is
not to say that a small fraction of absorbers do not exhibit more
extreme properties.  These extreme cases open the door to
some innovative analyses, permitting detection of unusual
species and even shedding light on the galaxy's
physical properties.

\subsection{Diffuse Interstellar Bands}
\begin{figure}
\centerline{\psfig{file=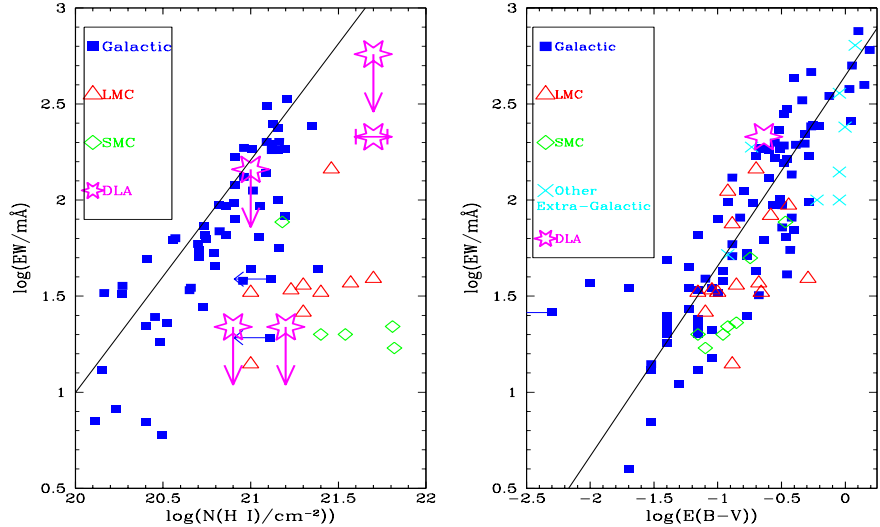,width=5.0in,height=3.0in,angle=270}}
\caption{A search for DIBs in DLAs; only one detection of the 5780 \AA\
DIB has been made (York et al. 2006).  In general the 5780 \AA\ DIB
is weaker for its N(HI) by up to a factor of 10 compared with Galactic
sightlines, but is consistent with strengths in the Magellanic Clouds.
In contrast, all sightlines show a similar dependence on E(B$-$V).
Figure adapted from York et al. (2006).}
\end{figure}

Diffuse interstellar bands (DIB) are common in the spectra of reddened
stars in the Milky Way.  Although numerous (there are now over 100
different known bands), the identification of the DIB carriers
is one of the oldest outstanding mysteries in astronomical
spectroscopy.  Amongst the potential
candidates are polycyclic aromatic hydrocarbons and long carbon
chains.  DIBs have been detected in only a handful of extra-galactic
sightlines, including the LMC and SMC (see Snow 2001 for a review)
and in one case, the broad 4428 \AA\ feature in a $z \sim 0.5$
DLA (Junkkarinen et al 2004). Since the strength of some DIB lines,
such as the 5780 \AA\ feature, correlate with N(HI) in the Galaxy (Herbig
1993) selecting 
high column density DLAs may allow us to detect DIBs in high redshift 
galaxies.  We have been undertaking such a search (Lawton
et al. in preparation) and summarise our findings in Figure 3.
In only one case do we detect DIB absorption: in the $z \sim 0.5$
DLA towards AO 0235+164 (York et al. 2006) where we detect
both the 5780 \AA\ and the 5705 \AA\ DIB.  From this detection
and upper limits from 4 other DLAs,
we find that the 5780 \AA\ line strengths in DLAs are weaker,
by up to a factor of 10, for their N(HI) than Galactic
sightlines.  Similar deficiencies are seen in Magellanic sightlines,
suggesting that DIB strength likely depends on metallicity
and local physical properties as well as N(HI). On the other hand,
the 5780 and 5705 \AA\ DIBs have similar equivalent width ratios
in the $z \sim 0.5$ DLA and the Galaxy, possible evidence that
these two bands originate from a similar character.  Surprisingly,
we do not detect the 6284 \AA\ DIB which is usually much
stronger than the 5780 \AA\ line which we do detect.  This indicates
that the physical conditions in the DLA are different to the
bulk of Galactic and Magellanic sightlines that have been
studied.  The only
known sightline with similarly weak 6284 \AA\ absorption is
towards the sightline Sk 143, located in the SMC wing.
Finally, in contrast to the trends with N(HI),
Galactic, Magellanic and DLA sightlines show a similar trend
of E(B$-$V) with DIB strength.  Since DLAs generally have low
E(B$-$V) (Ellison, Hall \& Lira 2005) this implies that DIBs are
unlikely to be commonly detected in these absorbers.

\subsection{Metal-Strong DLAs in the SDSS}

Herbert-Fort et al. (2006) have recently identified a sample of
`metal-strong' absorbers from the SDSS, characterised by
strong heavy element absorption that is clearly detected even
in low resolution SDSS spectra.  These metal-strong DLAs comprise
$\sim$ 5\% of the total population.  In some cases, the very strong
metal lines are a symptom of high N(HI).  In other cases, the
metallicity is truly high compared with known DLAs.  Although
relatively rare, the statistical power of Sloan is expected to
reveal several hundred metal-strong DLAs.  Follow-up observations
with high resolution spectrographs will yield data for
a myriad of applications, including the search for rarely detected
atomic transitions, molecular species and the study of isotopic ratios.

\medskip

In closing, it is interesting to note that although the definition
of `metal-rich' at this conference has been somewhat subjective,
with the exception of AGN (see e.g. Max Pettini's contribution
to these proceedings) we rarely find metallicities above $2 Z_{\odot}$
in either individual stars, HII regions or in galaxies.  However, a handful
of very super-solar abundances (up to $5 Z_{\odot}$) have been recently 
reported for QSO absorbers just below the traditional DLA
column density criterion of N(HI) $\ge 2 \times 10^{20}$ cm$^{-2}$
(P\'eroux et al 2006; Prochaska et al. 2006).  The surprisingly high 
metallicities in
these absorbers leads to a volley of new questions:  How have these
absorbers become so metal-rich at early times?  What are their
low redshift analogues? And what is the implication for the
cosmic metals budget?  

\acknowledgements

I am fortunate to enjoy stimulating collaborations with a large number of
of colleagues. In particular for the work described here, Chris Akerman, 
Chris Churchill, Pat Hall, Stephane Herbert-Fort, Lisa Kewley, Brandon Lawton,
Paulina Lira, Gabriela Mallen-Ornelas,
Max Pettini, Jason X. Prochaska, Ted Snow and Brian York.

\end{document}